\documentclass[aps,prl,8pt,twocolumn,showpacs,amsmath,amssymb,floatfix,footinbib,superscriptaddress]{revtex4-2}
\usepackage{mathtools,amsmath,amssymb,natbib,amsmath,mathrsfs,graphicx,graphics,geometry,CJK,color,overpic,subfig,subcaption,caption}
\usepackage{multirow,color,xcolor,bm,tabularx,psfrag,dcolumn,datetime,fancyhdr,float,comment}
\usepackage[bookmarksopen,bookmarksnumbered,colorlinks=true,citecolor=blue, linkcolor=blue, urlcolor=blue]{hyperref}
\usepackage{microtype}        
\usepackage{ragged2e}         
\usepackage{booktabs}         

\usepackage[pagewise,switch]{lineno}
\def\footnoterule{\kern -1mm \hrule width 6.6cm \kern 2.2mm}%
\usepackage{tikz,xcolor,hyperref}
\definecolor{lime}{HTML}{A6CE39}
\DeclareRobustCommand{\orcidicon}{%
    \begin{tikzpicture}
    \draw[lime, fill=lime] (0,0)
    circle [radius=0.16] node[white]
   {{\fontfamily{qag}\selectfont \tiny ID}};\draw[white, fill=white] (-0.0625,0.095)
    circle [radius=0.007];
    \end{tikzpicture}
    \hspace{-2mm}}
\foreach \x in {A, ..., Z}
{\expandafter\xdef\csname orcid\x\endcsname{\noexpand\href{https://orcid.org/\csname orcidauthor\x\endcsname}{\noexpand\orcidicon}}}

\begin{document}

\title{Charging Dynamics in a Distance-Modulated Planar Quantum-Battery Architecture}


\author{Yi-Fan Yang}
\affiliation{Center for Quantum Materials and Computational Condensed Matter Physics, Faculty of Science, Kunming University of Science and Technology, Kunming, 650500, PR China}
\affiliation{Department of Physics, Faculty of Science, Kunming University of Science and Technology, Kunming, 650500, PR China}
\author{Shun-Cai Zhao\orcidA{}}
\email[Corresponding author: ]{zsczhao@126.com.}
\affiliation{Center for Quantum Materials and Computational Condensed Matter Physics, Faculty of Science, Kunming University of Science and Technology, Kunming, 650500, PR China}
\affiliation{Department of Physics, Faculty of Science, Kunming University of Science and Technology, Kunming, 650500, PR China}

\begin{abstract}
While the spatial arrangement of individual units is essential for the physical implementation of quantum batteries, geometry-dependent interactions are rarely explicitly incorporated into existing theoretical models. To address this, we propose a planar many-body quantum-battery architecture consisting of coupled resonators. By introducing a distance-dependent function to modulate both the inter-battery coupling and tunneling, we investigate the open-system charging dynamics in the strong-coupling regime using a Redfield master-equation approach. Using ergotropy as the primary figure of merit, we demonstrate that the charging performance is highly sensitive to the inter-battery distance, nearest-neighbor coupling strength, and environmental conditions. Specifically, decreasing the inter-battery distance within an optimal window suppresses charging fluctuations and accelerates the system's approach to a steady charged state. However, an excessively short distance amplifies environmental dissipation, thereby degrading the overall performance. Furthermore, while overly strong inter-battery coupling induces post-charging instability, moderate coupling achieves a favorable balance between maximum stored energy and stability. We also establish that the system-bath coupling and bath cutoff frequency predominantly govern the charging timescale, and that the planar architecture maintains its robustness against thermal fluctuations over a broad temperature range. These results highlight the critical role of geometry-controlled interactions in many-body quantum batteries, providing a theoretical foundation for the design and optimization of two-dimensional quantum energy-storage devices.
\end{abstract}

\maketitle
\section{Introduction}\label{INTRODUCTION}

Quantum batteries (QBs) have emerged as a promising framework for investigating energy storages~\cite{assawaworrarit_robust_2017} and extraction at the quantum scale, particularly in quantum thermodynamics and nonequilibrium many-body physics. In contrast to classical energy-storage devices, QBs may exploit genuinely quantum features\cite{3tm5-vsqw,PhysRevE.103.042118,daley_practical_2022,PhysRevA.107.032203}---including coherence\cite{PhysRevLett.126.220404,PhysRevA.100.052311,fan_quantifying_2022}, correlations\cite{PhysRevResearch.3.033135,Paneru_2020,zhao_entanglement_2025,kauffman_topological_2019}, and collective effects cite{PhysRevLett.132.140802,PhysRevLett.117.250503}---to enhance charging performance or extractable work\cite{PhysRevApplied.21.040501,PhysRevA.99.022314,zhu_interactive_2023}. This possibility has motivated extensive theoretical research on charging power, ergotropy, and quantum advantage in both few-body and many-body settings.

In recent years, research on QBs has mainly developed along three directions. First, single- or few-body models~\cite{PhysRevA.108.052213,PhysRevA.109.052206,PhysRevLett.132.210402}have established foundational concepts regarding charging power and extractable work. The second direction focuses on collective charging and many-body quantum advantages, including entanglement-assisted charging\cite{PhysRevA.104.L030402,PhysRevB.104.245418}, superabsorption-enhanced schemes\cite{Superabsorption_quantum_battery}, and one-dimensional many-body architectures such as spin-chain QBs\cite{PhysRevA.103.033715,PhysRevA.106.032212,evangelakos_fast_2025,PhysRevA.97.022106}. The third direction moves toward physically motivated implementations and prototype-oriented platforms, including microcavity-based QB models in the photovoltaic setting\cite{James_2022} and QB-inspired scenarios relevant to fast charging and high-density energy storage\cite{PhysRevResearch.6.023136,PhysRevA.104.032207,romero_scrambling_2025}. Collectively, these studies have significantly advanced the theoretical foundations of QBs and clarified a variety of mechanisms for quantum-enhanced charging.

However, existing theoretical models often lack realistic spatial architectures. While 1D and few-body systems demonstrate cooperative speedups, they typically assume idealized, geometry-independent couplings. In fabricated QB arrays, spatial arrangement inherently dictates interaction strengths and energy pathways. Consequently, neglecting geometry can yield misleading predictions regarding charging stability, useful energy storage, and dissipation. This geometric dependence is particularly critical in open quantum systems, where environmental noise and dissipation intertwine with spatial layout. For example, reducing inter-battery distance strengthens coherent transport but risks amplifying environmental losses and post-charging instability. Realistic charging therefore emerges from a nontrivial competition between geometry-controlled interactions and dissipation. Understanding this interplay is essential to determine if theoretical quantum advantages persist in structured, dissipative architectures~\cite{8xsm-5mb6}.

In this work, we propose a planar many-body QB composed of coupled resonators, using a distance-dependent function to encode spatial arrangement as a genuine control parameter. Operating in the strong-coupling regime, we model the open-system dynamics via the Redfield master equation and evaluate the work extraction potential via ergotropy. Our aim is to identify how geometric control and environmental parameters jointly affect the charging timescale, the attainable useful energy, and the temporal stability of the charged state. The architecture exhibits remarkable robustness against thermal fluctuations, with the global charging timescale predominantly governed by the system-bath coupling and the spectral cutoff frequency. 

The remainder of this paper is organized as follows. Sec.~II introduces the planar QB model and its distance-dependent coupling. Sec.~III details the open-system formalism and performance metrics. Sec.~IV presents numerical results under varying geometric and environmental conditions. Sec.~V concludes with a summary and future outlook.

\section{Model of the planar quantum batteries}\label{MODEL}
 
 \begin{figure}[tb]
  \centering
  \includegraphics[width=0.35\textwidth]{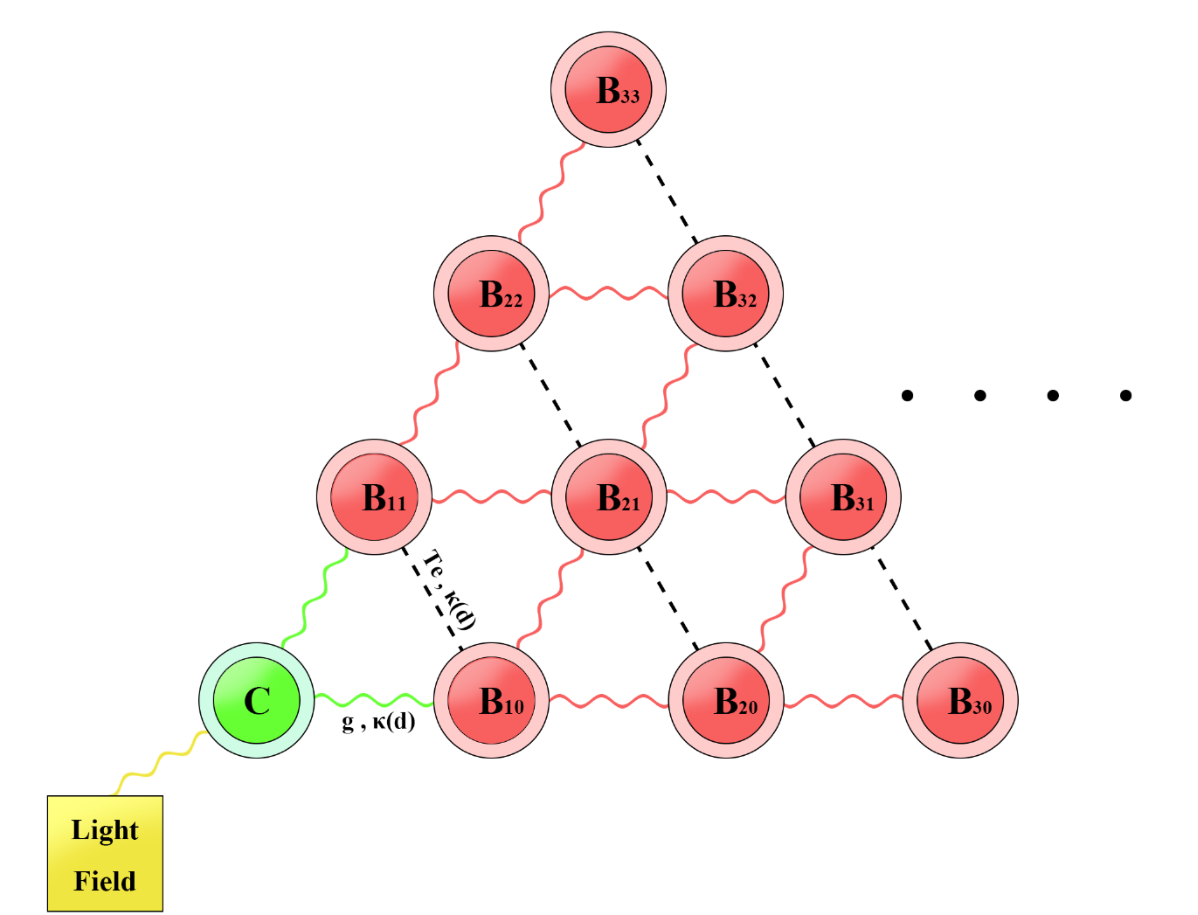}
  \caption{Schematic illustration of the charging architecture for a planar many-body quantum battery (QB) array. 
    The system consists of a central charger $C$ and multiple layers of QB cells ($B_{ij}$), modeled as harmonic resonators. 
    An external coherent field drives the charger, initiating a cascaded energy transfer through the array. 
    The inter-layer coupling $g$ and intra-layer tunneling $T_e$ (indicated by dashed lines) are both modulated by a distance-dependent scaling factor $\kappa(d)$. 
    This configuration enables collective charging of the planar array via the competition between mediated interactions and spatial-dependent attenuation.
}\label{fig1}
\end{figure}

We consider a planar quantum-battery architecture composed of a central charger $C$ and $n$ layers of coupled harmonic-oscillator battery cells arranged in a two-dimensional geometry, as illustrated in Fig.~\ref{fig1}. The total Hamiltonian is written as
\begin{equation}\label{Eq1}
\hat{H}_{total}=\hat{H}_c+\hat{H}_s+\hat{H}_E+\hat{H}_I,
\end{equation}
where $\hat{H}_{\mathrm{c}}$ describes the charger, $\hat{H}_{\mathrm{s}}$ the battery array and its coherent driving, $\hat{H}_{\mathrm{E}}$ the environmental bath, and $\hat{H}_{\mathrm{I}}$ the system-bath interaction.

The charger  $C$  is modeled as a harmonic mode,
\begin{equation}\label{Eq2}
\hat{H}_c=\omega_c \hat{a}^\dagger \hat{a} .
\end{equation} 

The battery array consists of harmonic cells $\hat{b}_{ij}$, where the index $i$ labels the layer ( $i = 1,2,\dots,n$) and $j$ labels the cell within that layer($j = 0,1,\dots,i$). To account for the geometric arrangement of the array, we introduce a distance-dependent modulation factor $\kappa(d)$, which rescales both the charger-battery coupling and the tunneling processes between neighboring cells. 
Therefore, the battery array and its coherent driving Hamiltonian $\hat{H}_{\mathrm{s}}$ contains five ingredients: the free energy of the battery cells, coherent energy injection from the charger, coherent energy injection from inter-layer coupling, intra-layer tunneling transport inside the planar array, and an external coherent drive applied to the charger, as follows,

\begin{align}\label{Eq3}
\hat{H}_s &= \sum_{i=1}^{n} \sum_{j=0}^{i} \omega_{ij} \hat{b}_{ij}^\dagger \hat{b}_{ij} \nonumber\\
&+ \kappa(d) \left[ g_{0,10} \hat{a}^\dagger \hat{b}_{10} + g_{0,11} \hat{a}^\dagger \hat{b}_{11} + \text{H.c.} \right] \nonumber\\
&+\kappa(d) \sum_{i=1}^{n-1} \sum_{j=0}^i \resizebox{.75\columnwidth}{!}{$\left[ g_{ij,(i+1)j} \hat{b}_{ij}^\dagger \hat{b}_{(i+1)j}  + g_{ij,(i+1)(j+1)} \hat{b}_{ij}^\dagger \hat{b}_{(i+1)(j+1)} + \text{H.c.} \right]$} \nonumber\\
&+ \kappa(d) \sum_{i=1}^n \sum_{j=0}^{i-1} T_e \left[ \hat{b}_{ij}^\dagger \hat{b}_{i(j+1)} + \text{H.c.} \right] \nonumber\\
&+ F \left( \hat{a} e^{i\omega_f t} + \hat{a}^\dagger e^{-i\omega_f t} \right).
\end{align}

Here, the first term accounts for the free Hamiltonian of the QB cells, with each cell modeled as an independent harmonic oscillator with frequency $\omega_{ij}$. The second term characterizes the interface between the charger $C$ and the first battery layer, where the coupling is governed by the strengths $g_{0,ij}$ and a distance-dependent scaling factor $\kappa(d)$. The subsequent terms describe the connectivity within the array: inter-layer interactions facilitate energy transport along the stacking direction (the charger axis), while horizontal intra-layer terms represent tunneling between adjacent cells within the same layer with amplitude $T_e$. Considering the isotropy of energy transport, the identical distance-dependent function $\kappa(d)$ is introduced along the inter-layer direction (the stacking direction, i.e., the charger axis) and within the horizontal intra-layer between adjacent batteries to modulate both the inter-battery coupling and tunneling. Finally, the system is driven by an external optical field with amplitude $F$ and frequency $\omega_f$. Note that for all interaction terms, $\text{H.c.}$ denotes the Hermitian conjugate. 

A key feature of the present model is that geometry enters the Hamiltonian explicitly through $\kappa(d)$. In this way, the intercell distance is promoted from a purely structural parameter to a dynamical control parameter that directly modifies the charging pathways. This construction is intended as a phenomenological description of planar architectures in which the coherent couplings depend on spatial separation, as is commonly encountered in circuit-QED and atomic-array settings.

The charging dynamics follow a cascaded energy transfer mechanism initiated by the external optical drive. Initially, the charger $C$ is energized by the external field and subsequently couples to the first layer of batteries ($B_{10}$ and $B_{11}$) with a strength characterized by $\kappa(d) g_{0,ij}$. As these initial cells are populated, energy is concurrently transferred to the second layer ($B_{20}$, $B_{21}$, and $B_{22}$) and beyond. This sequential process is mediated by inter-layer interactions ( $\kappa(d) g_{ij,(i+1)j}$, $\kappa(d) g_{ij,(i+1)(j+1)}$) and intra-layer tunneling, with the parameter $\kappa(d) T_e$. Ultimately, this cascaded coupling scheme enables the collective charging of the entire planar quantum battery array, as schematically illustrated in Fig.~\ref{fig1}.

In Eq.~\eqref{Eq1}, the environment is modeled as a multimode bosonic bath,
\begin{equation}\label{Eq4}
\hat{H}_E = \sum_k \omega_k \hat{c}_k^\dagger \hat{c}_k,
\end{equation}
where $\omega_k$ are the frequencies of the individual modes, and $\hat{c}_k^\dagger$ ($\hat{c}_k$) are the corresponding creation (annihilation) operators. 
The interaction between the planar battery and the environment is taken in the standard bilinear form
\begin{equation}\label{Eq5}
\hat{H}_I = \sum_{i=0}^{n} \sum_k G_{i,k} \,\hat{B}_i \left( \hat{c}_k + \hat{c}_k^\dagger \right),
\end{equation}
where $\hat{B}_i$ are system operators (e.g., level-projectors or hopping operators) associated with the $i$th site or level of the battery, and $G_{i,k}$ are real coupling strengths. The form $\hat{c}_k + \hat{c}_k^\dagger$ corresponds to a position-like coupling, which is standard for linear system-bath interactions. In the following, we work within the Redfield-type master equation\cite{PhysRevA.109.052205,PhysRevLett.129.200403,PhysRevE.94.022126} for the battery's reduced density operator.

\begin{eqnarray}\label{Eq6}
\begin{split}
\frac{d{\hat{\rho}}(t)}{dt}=&-\frac{i}{\hbar}[\hat{H}_s, \hat{\rho}(t)]\\
        &+\sum_{ij}R_{ij}[2\hat{L}_{ij}\hat{\rho}(t)\hat{L}^\dagger_{ij}-\hat{L}^\dagger_{ij}\hat{L}_{ij}\hat{\rho}(t)],
\end{split}
\end{eqnarray}
where \(R_{ij}\) represents the Redfield tensor that governs energy dissipation to the bath during the transition process. The Redfield tensor\cite{R_Kubo_1966} is a quantity dependent on the bath spectral density and temperature, and the bath spectral density, and its form can be expressed as follows:

\begin{eqnarray}\label{Eq7}
\begin{split}
R_{ij}=J_{ij}[\coth(\frac{\hbar\omega_k}{2k_BT})+1],
\end{split}
\end{eqnarray}

\noindent where \(J_{ij}\) describes the spectral density of the coupling between the bath and the system, thereby indirectly regulating the dissipation behavior between the system and the bath. Based on the bath Hamiltonian presented earlier, we adopt a Debye-type form, given by:

\begin{eqnarray}\label{Eq8}
\begin{split}
J_{ij}=\gamma_{ij}\cdot\frac{\omega_k}{\omega^2_0+\omega_k^2},
\end{split}
\end{eqnarray}

\noindent where \(\gamma_{ij}\)represents the coupling coefficient between the system and the environment, \(\omega_k\) denotes the bath frequency and \(\omega_0\) is the cutoff frequency. 

While \(L_{ij}\) in \eqref{Eq6}, defined as the Lindblad-like\cite{10.1093/acprof:oso/9780199213900.001.0001} transition operator, connects the eigenstates of the system Hamiltonian \(\hat{H}_s\).
In the numerical simulations, we solve the Redfield equation using the time-discretization method for the numerical solution. The ergotropy $\mathcal{E}(t)$is then computed \cite{Allahverdyan2004} as
\begin{align}\label{Eq9}
\mathcal{E}(t) = \text{Tr}\left[ \hat{\rho}(t) \hat{H}_s \right] - \min_{\hat{U}} \text{Tr}\left[ \hat{U} \hat{\rho}(t) \hat{U}^\dagger \hat{H}_s \right],
\end{align}
where the minimization is over all unitary transformations $\hat{U}$, which measures the maximum work extractable from the instantaneous battery state by unitary operations..

For the numerical analysis, we focus on the minimal planar charging cell formed by the charger and its nearest battery units, namely the first triangular sector of the array. This reduced sector is not intended to replace the full planar architecture; rather, it provides the minimal local module in which the elementary charging process can be resolved most transparently. Such a restriction is well motivated by our emphasis on the initial stage of charging, where the external drive first populates the charger and the injected energy is transferred predominantly to the nearest battery cells. In this short-time regime, the dynamics is governed primarily by the local coherent scale $g\kappa(d)$ and by the intra-array tunneling amplitude $T_e$, which controls the redistribution of energy within the first layer, whereas more distant cells mainly contribute to later-time cascaded transport and additional collective effects. 

In this sense, the reduced model captures the essential local physics of coherent injection, intralayer energy sharing, and distance-dependent attenuation while keeping the analysis analytically and numerically transparent. The full planar many-body battery may then be viewed as an extended assembly of such local charging units, supplemented by interlayer propagation processes beyond the present minimal description.

\section{Results and discussions}\label{Results and disscussions}

Using the reduced planar charging cell defined in Sec.~II as the basic local module, we now examine the charging dynamics of the system through the time evolution of the ergotropy. Our analysis focuses on how the distance-dependent coupling factor $\kappa(d)$, the coherent parameters $g$ and $T_e$, and the environmental quantities controlling dissipation jointly affect the charging timescale, the attainable useful energy, and the stability of the charged state.

Accordingly, Eq.~\eqref{Eq3} can be explicitly written as
\begin{align}\label{Eq10}
\hat{H}_s
&=
\omega_{10}\hat{b}_{10}^\dagger \hat{b}_{10}
+\omega_{11}\hat{b}_{11}^\dagger \hat{b}_{11}
\nonumber\\[4pt]
&\quad
+ \kappa(d)\Big[
g_{0,10}\hat{a}^\dagger \hat{b}_{10}
+ g_{0,11}\hat{a}^\dagger \hat{b}_{11}
+ \text{H.c.}
\Big]
\nonumber\\[4pt]
&\quad
+ \kappa(d) T_e \Big[
\hat{b}_{10}^\dagger \hat{b}_{11}
+ \text{H.c.}
\Big]
\nonumber\\[4pt]
&\quad
+ F\left(
\hat{a} e^{i\omega_f t}
+ \hat{a}^\dagger e^{-i\omega_f t}
\right).
\end{align}

To model an isotropic planar many-body quantum battery, we take in Eq.~\eqref{Eq10} all harmonic oscillators(HO) constituting the battery units to have the same frequency $\omega$, and all coupling coefficients $g$ to be equal. The distance-dependent scaling factor $\kappa(d)$ is introduced as a dimensionless function of the interlayer or intralayer separation between different HO units \cite{Scali02012020,PhysRevLett.91.267401,PhysRevB.98.155320,Zoubi_2010}. Here $d=s/\omega$ denotes the dimensionless distance, and we take $\kappa(d)=\exp(-d)$. The corresponding parameter values are listed in Tab.~~\ref{tab1}.

\begin{table}[tb]  
    \centering     
    \caption{Parameters used in the numerical simulations for the HO battery unit}  
    \begin{tabular}{c|c|c|c|c|c}  
        \noalign{\hrule height 1pt}  
                   & $\omega(Hz)$ & $g(Hz)$      & $\gamma\times10^{-6}(Hz)$ & $\omega_0(Hz)$ & $T(K)$ \\
        \noalign{\hrule height 0.4pt} 
        Fig.2(a)   & $\backslash$ & 0.01         &           1               & 0.05           & 250\\
        Fig.2(b)   & 4.00         & $\backslash$ &           1               & 0.05           & 300\\
        Fig.2(c)   & 3.00         & 0.01         &           1               & 0.05           & 300\\
        Fig.3(a)   & 3.00         & 0.01         &      $\backslash$         & 0.05           & 300\\
        Fig.3(b)   & 4.00         & 0.01         &           1               & $\backslash$   & 300\\
        Fig.3(c)   & 4.00         & 0.01         &           1               & 0.05           & $\backslash$\\
        \noalign{\hrule height 1pt} 
    \end{tabular}
    \label{tab1}  
\end{table}

\subsection{Dependence on geometric parameters}

At the outset of the following discussion, it is worth emphasizing that the charging performance of the present planar many-body quantum battery is governed mainly by two effective quantities, namely the interlayer coupling $g\kappa(d)$ and the intralayer tunneling $\kappa(d)T_e$. Since these two terms directly determine the effective coupling architecture and the associated energy-transport channels, they jointly control the ergotropy. In what follows, we analyze the dynamical behavior of the ergotropy by examining the respective roles of the distance $d$, the coupling strength $g$, and the tunneling amplitude $T_e$.
\begin{figure}[tp]
        \centering
        \subfloat
        {
        \begin{overpic}[width=0.4\textwidth]{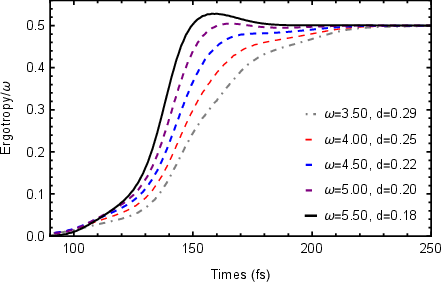}
        \put(13,15){\scalebox{0.9}{(a)}} 
        \end{overpic}
        \captionsetup{labelformat=empty}
        \label{fig2a}               
        }
        
        \subfloat
        {
        \begin{overpic}[width=0.4\textwidth]{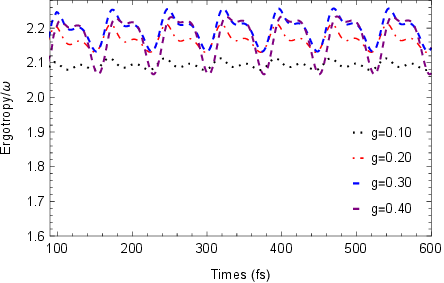}
        \put(14,14){\scalebox{0.9}{(b)}}
        \end{overpic}   
        \captionsetup{labelformat=empty}
        \label{fig2b}              
        }
        
        \subfloat
        {
        \begin{overpic}[width=0.38\textwidth]{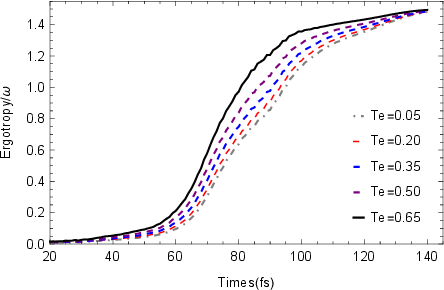}
        \put(14,14){\scalebox{0.9}{(c)}}
        \end{overpic}   
        \captionsetup{labelformat=empty}
        \label{fig2c}              
        }    
        \caption{Time-dependent evolution of the ergotropy of QB unit obtained by varying the parameters of the QB system. (a) Inter-layer distance $d$ with $T_e=0.001$ Hz; (b) Inter-layer coupling coefficient $g$ with $T_e=0.001$ Hz; (c) Intra-layer tunneling coefficient $T_e$. Here, $\gamma_{ij}=\gamma$, $s$=1, $\omega_k=0.085$ Hz. All other parameters are taken from Tab.~\ref{tab1}.   }
        \label{fig2}
\end{figure}

Fig.~\ref{fig2} illustrates the time evolution of the ergotropy under the variation of three key structural parameters: the distance $d$, the interlayer coupling $g$, and the intralayer tunneling $T_e$. As shown in Fig.~\ref{fig2}(a), increasing $d$ significantly delays the charging process without altering the achievable peak value of the ergotropy. In contrast, enhancing the interlayer coupling $g$ [Fig.~\ref{fig2}(b)] increases the peak ergotropy but simultaneously amplifies temporal oscillations, rendering the energy output less stable. Finally, Fig.~\ref{fig2}(c) demonstrates that the ergotropy peak grows monotonically with the intralayer tunneling $T_e$.

The physical mechanisms underlying these distinct dynamical behaviors can be naturally understood through the interplay of the two effective energy scales governing the local sector: the injection rate $g\kappa(d)$ and the redistribution rate $\kappa(d)T_e$. 
Specifically, the distance $d$ enters the dynamics solely through the global scaling factor $\kappa(d)$, proportionally renormalizing both the charger-battery coupling and the intralayer tunneling. Consequently, varying $d$ primarily rescales the overall charging timescale. A larger distance weakens the effective transport rates, thereby prolonging the time required to reach the maximum ergotropy, while leaving the ultimate storage capacity unaffected in this local short-time regime.

Conversely, the parameter $g$ directly dictates the coherent energy-injection channel from the charger to the first battery layer. A larger $g$ strengthens the coherent pumping, which naturally yields a higher ergotropy peak. However, this stronger hybridization also intensifies the reversible energy exchange (i.e., Rabi-like back-and-forth oscillations) between the charger and the battery cells. As a result, while beneficial for the absolute charging capacity, a strong $g$ inevitably amplifies temporal fluctuations and compromises the charging stability.

Finally, $T_e$ governs the intralayer transport channel, determining the efficiency of energy redistribution within the battery layer. For small $T_e$, injected excitations remain highly localized, creating a local bottleneck that limits cooperative storage. Increasing $T_e$ facilitates rapid energy spreading among neighboring cells, thereby mitigating local saturation. This spatial sharing of excitations enables a more efficient collective storage, translating into a larger amount of extractable work.

In summary, these three parameters play distinct yet complementary roles in shaping the battery performance: $d$ dictates the charging timescale, $g$ controls the injection strength and coherent oscillations, and $T_e$ governs the efficiency of local spatial sharing. Their combined optimization is therefore essential for achieving both high capacity and temporal stability in planar quantum batteries.

\begin{figure}[tp]
        \centering
        \subfloat
        {
        \begin{overpic}[width=0.4\textwidth]{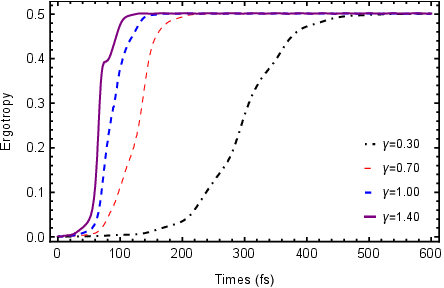}
        \put(13,14){\scalebox{0.9}{(a)}} 
        \end{overpic}
        \captionsetup{labelformat=empty}
        \label{fig3a}               
        }
        
        \subfloat
        {
        \begin{overpic}[width=0.4\textwidth]{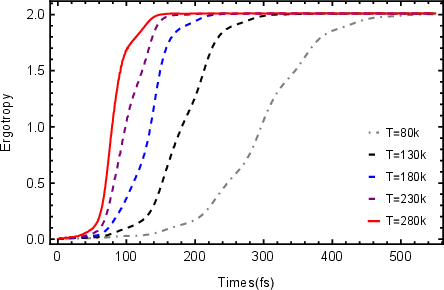}
        \put(13,14){\scalebox{0.9}{(b)}}
        \end{overpic}   
        \captionsetup{labelformat=empty}
        \label{fig3b}              
        }
        
          \subfloat
        {
        \begin{overpic}[width=0.4\textwidth]{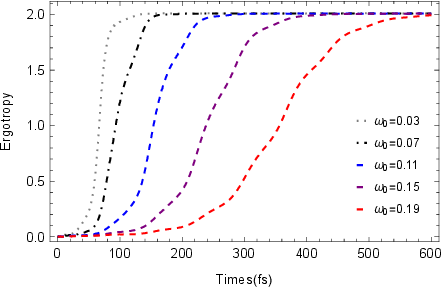}
        \put(13,14){\scalebox{0.9}{(c)}}
        \end{overpic}   
        \captionsetup{labelformat=empty}
        \label{fig3c}              
        }
        \caption{
        Time-dependent evolution of the ergotropy of QB unit obtained by varying the environmental parameters. (a) System-environment coupling coefficient \(\gamma\); (b) Environment temperature \(T\); (c) Cutoff frequency \(\omega_0\). All other parameters are the same to those in Fig.~\eqref{fig2}.
        }
        \label{fig3}
\end{figure}

\subsection{Dependence on environmental parameters}

Building upon this open-system framework, Fig.~\ref{fig3} illustrates the time evolution of the ergotropy under the modulation of various environmental parameters. We can clearly identify two distinct types of dynamical control: acceleration and retardation of the charging stabilization. As shown in Figs.~\ref{fig3a} and \ref{fig3b} , increasing either the system-environment coupling strength $\gamma$ or the bath temperature $T$ yields a constructive effect. Under these conditions, the time required to reach the stable peak of the ergotropy is substantially shortened, while the maximum accessible ergotropy itself remains robustly unchanged. Such behavior implies that a stronger or hotter environment can facilitate a rapid and stable energy output. Conversely, Fig.~\ref{fig3c} reveals a detrimental effect associated with the bath cutoff frequency: increasing $\omega_0$ notably prolongs the transient charging regime, thereby delaying the moment at which the battery reaches its optimal stable capacity.

The underlying physical mechanisms governing these acceleration and delay behaviors can be systematically understood by analyzing the Redfield tensor $R_{ij}$, which dictates the characteristic timescale of energy dissipation and decoherence. In our driven-dissipative battery setup, the long-time stable output corresponds to a non-equilibrium steady state, which is established through the competition between the coherent driving field and environmental dissipation. Consequently, any parameter variation that enhances the dissipation rate $R_{ij}$ will accelerate the damping of the transient Rabi-like oscillations, forcing the system to converge to its stationary pumped state more rapidly.

Specifically, as dictated by Eq.~\eqref{Eq8}, the system-environment coupling $\gamma$ linearly scales the spectral density $J_{ij}$. Thus, a larger $\gamma$ directly amplifies the relaxation rate $R_{ij}$, leading to the accelerated stabilization observed in Fig.~\ref{fig3}(a). Similarly, a higher bath temperature $T$ increases the thermal phonon occupation. This amplifies the thermal factor $\coth(\hbar\omega_k / 2k_B T)$ in Eq.~\eqref{Eq7}, resulting in a phonon-assisted enhancement of the dissipation tensor $R_{ij}$. Therefore, elevating the temperature effectively acts as a thermal catalyst that fast-tracks the charging process without degrading the final stored energy [Fig.~\ref{fig3}(b)].

On the other hand, the cutoff frequency $\omega_0$ behaves entirely differently because it appears in the denominator of the Debye spectral density [Eq.~\eqref{Eq8}]. Increasing $\omega_0$ effectively suppresses the spectral weight of the system-bath coupling at the relevant transition frequencies, yielding a smaller $J_{ij}$ and, correspondingly, a reduced dissipation rate $R_{ij}$. This weakened environmental damping means the system requires a much longer time to wash out the transient fluctuations and settle into the steady state, which perfectly explains the delayed stabilization phenomenon observed in Fig.~\ref{fig3}(c). 

In short, the environmental parameters shape the battery's performance by tuning the effective relaxation timescale: variables that enhance the dissipation rate ($\gamma$ and $T$) accelerate the stabilization of the output, whereas those that suppress dissipation ($\omega_0$) induce a temporal delay.

\section{Experimental feasibility and outlook}\label{POSSIBLE EXPERIMENTAL REALIZATION}

The present model is relevant to experimental platforms in which coherent couplings can be engineered through spatial layout, most notably superconducting circuit-QED arrays and Rydberg-atom lattices. In circuit-QED implementations, the charger and battery units may be realized by microwave resonators or transmon-like modes, with the charger-battery coupling $g$ and the effective intercell tunneling $T_e$ controlled by capacitive or inductive circuit elements. The distance-dependent factor $\kappa(d)$ can then be viewed as an effective parametrization of geometry-dependent couplings in planar layouts. In Rydberg arrays, by contrast, the spatial dependence is more direct, since the interaction strength is naturally tunable through the interatomic separation. The external driving parameters $F$ and $\omega_f$ correspond to standard coherent microwave or laser drives, while the dissipative quantities $\gamma$, $\omega_0$, and $T$ represent the effective relaxation rate, bath spectral scale, and environmental temperature, respectively.

At the same time, the present description remains idealized. A more realistic implementation should account for parameter inhomogeneity, long-range cross talk, non-Markovian effects, and finite-size boundary corrections in extended two-dimensional arrays. In addition, although ergotropy is a useful theoretical figure of merit, its experimental extraction would generally require state reconstruction or related indirect protocols. For this reason, the current results should be viewed as identifying a feasible design principle rather than a complete hardware-level proposal. A natural next step is to combine the present geometry-dependent charging mechanism with platform-specific dissipative modeling in order to assess larger planar arrays under realistic experimental conditions.

\section{Conclusion}\label{CONCLUSION}

In conclusion, we have investigated the non-equilibrium charging dynamics of a distance-modulated planar quantum battery within a driven-dissipative framework. By explicitly embedding spatial scaling factors into the many-body Hamiltonian, we demonstrated that spatial layout acts as an active control parameter rather than a passive structural feature, collectively dictating the charging capacity and the stabilization timescale of the system.

Our numerical analysis reveals a clear interplay between geometric arrangement and environmental effects. Specifically, scaling down the intercell distance $d$ accelerates the charging rate, provided it remains above a critical threshold to avoid excessive spatially-induced dissipation. Furthermore, we find that the planar architecture exhibits notable robustness against environmental noise. Counterintuitively, strengthening the system-environment coupling $\gamma$ and operating at elevated temperatures $T$ can act to accelerate the convergence to the steady-state maximum ergotropy, while reducing the environment cutoff frequency $\omega_0$ effectively suppresses the retardation of the charging process.

While this distance-dependent framework offers a physically motivated step beyond idealized one-dimensional models, we emphasize that the present analysis is restricted to a minimal planar charging unit and a perturbative dissipative treatment. To fully assess the collective many-body advantages and experimental viability in platforms such as superconducting grids or Rydberg arrays, future studies must extend this theoretical framework to larger two-dimensional lattices and incorporate platform-specific hardware constraints. Within these bounds, our results provide a robust foundation for geometry-aware design in open quantum energy storage devices.

\section*{Author contributions}
S. C. Zhao conceived the idea. Y. F. Yang performed the numerical computations and wrote the draft, and S. C. Zhao did the analysis and revised the paper.
\bibliographystyle{plain}  
\bibliography{references}  
\bibliographystyle{apsrev4-2}

\end{document}